\documentclass[twocolumn,showpacs,preprintnumbers,amsmath,amssymb,superscriptaddress]{revtex4}

\usepackage{graphicx}
\usepackage{dcolumn}
\usepackage{bm}

\begin{document}


\preprint{APS/123-QED}

\title{Limits on a muon flux from Kaluza-Klein dark matter annihilations in the Sun \\ from the IceCube 22-string detector}

\affiliation{III. Physikalisches Institut, RWTH Aachen University, D-52056 Aachen, Germany}
\affiliation{Dept.~of Physics and Astronomy, University of Alabama, Tuscaloosa, AL 35487, USA}
\affiliation{Dept.~of Physics and Astronomy, University of Alaska Anchorage, 3211 Providence Dr., Anchorage, AK 99508, USA}
\affiliation{CTSPS, Clark-Atlanta University, Atlanta, GA 30314, USA}
\affiliation{School of Physics and Center for Relativistic Astrophysics, Georgia Institute of Technology, Atlanta, GA 30332. USA}
\affiliation{Dept.~of Physics, Southern University, Baton Rouge, LA 70813, USA}
\affiliation{Dept.~of Physics, University of California, Berkeley, CA 94720, USA}
\affiliation{Lawrence Berkeley National Laboratory, Berkeley, CA 94720, USA}
\affiliation{Institut f\"ur Physik, Humboldt-Universit\"at zu Berlin, D-12489 Berlin, Germany}
\affiliation{Fakult\"at f\"ur Physik \& Astronomie, Ruhr-Universit\"at Bochum, D-44780 Bochum, Germany}
\affiliation{Physikalisches Institut, Universit\"at Bonn, Nussallee 12, D-53115 Bonn, Germany}
\affiliation{Universit\'e Libre de Bruxelles, Science Faculty CP230, B-1050 Brussels, Belgium}
\affiliation{Vrije Universiteit Brussel, Dienst ELEM, B-1050 Brussels, Belgium}
\affiliation{Dept.~of Physics, Chiba University, Chiba 263-8522, Japan}
\affiliation{Dept.~of Physics and Astronomy, University of Canterbury, Private Bag 4800, Christchurch, New Zealand}
\affiliation{Dept.~of Physics, University of Maryland, College Park, MD 20742, USA}
\affiliation{Dept.~of Physics and Center for Cosmology and Astro-Particle Physics, Ohio State University, Columbus, OH 43210, USA}
\affiliation{Dept.~of Astronomy, Ohio State University, Columbus, OH 43210, USA}
\affiliation{Dept.~of Physics, TU Dortmund University, D-44221 Dortmund, Germany}
\affiliation{Dept.~of Subatomic and Radiation Physics, University of Gent, B-9000 Gent, Belgium}
\affiliation{Max-Planck-Institut f\"ur Kernphysik, D-69177 Heidelberg, Germany}
\affiliation{Dept.~of Physics and Astronomy, University of California, Irvine, CA 92697, USA}
\affiliation{Laboratory for High Energy Physics, \'Ecole Polytechnique F\'ed\'erale, CH-1015 Lausanne, Switzerland}
\affiliation{Dept.~of Physics and Astronomy, University of Kansas, Lawrence, KS 66045, USA}
\affiliation{Dept.~of Astronomy, University of Wisconsin, Madison, WI 53706, USA}
\affiliation{Dept.~of Physics, University of Wisconsin, Madison, WI 53706, USA}
\affiliation{Institute of Physics, University of Mainz, Staudinger Weg 7, D-55099 Mainz, Germany}
\affiliation{University of Mons-Hainaut, 7000 Mons, Belgium}
\affiliation{Bartol Research Institute and Department of Physics and Astronomy, University of Delaware, Newark, DE 19716, USA}
\affiliation{Dept.~of Physics, University of Oxford, 1 Keble Road, Oxford OX1 3NP, UK}
\affiliation{Dept.~of Physics, University of Wisconsin, River Falls, WI 54022, USA}
\affiliation{Oskar Klein Centre and Dept.~of Physics, Stockholm University, SE-10691 Stockholm, Sweden}
\affiliation{Dept.~of Astronomy and Astrophysics, Pennsylvania State University, University Park, PA 16802, USA}
\affiliation{Dept.~of Physics, Pennsylvania State University, University Park, PA 16802, USA}
\affiliation{Dept.~of Physics and Astronomy, Uppsala University, Box 516, S-75120 Uppsala, Sweden}
\affiliation{Dept.~of Physics and Astronomy, Utrecht University/SRON, NL-3584 CC Utrecht, The Netherlands}
\affiliation{Dept.~of Physics, University of Wuppertal, D-42119 Wuppertal, Germany}
\affiliation{DESY, D-15735 Zeuthen, Germany}

\author{R.~Abbasi}
\affiliation{Dept.~of Physics, University of Wisconsin, Madison, WI 53706, USA}
\author{Y.~Abdou}
\affiliation{Dept.~of Subatomic and Radiation Physics, University of Gent, B-9000 Gent, Belgium}
\author{T.~Abu-Zayyad}
\affiliation{Dept.~of Physics, University of Wisconsin, River Falls, WI 54022, USA}
\author{J.~Adams}
\affiliation{Dept.~of Physics and Astronomy, University of Canterbury, Private Bag 4800, Christchurch, New Zealand}
\author{J.~A.~Aguilar}
\affiliation{Dept.~of Physics, University of Wisconsin, Madison, WI 53706, USA}
\author{M.~Ahlers}
\affiliation{Dept.~of Physics, University of Oxford, 1 Keble Road, Oxford OX1 3NP, UK}
\author{K.~Andeen}
\affiliation{Dept.~of Physics, University of Wisconsin, Madison, WI 53706, USA}
\author{J.~Auffenberg}
\affiliation{Dept.~of Physics, University of Wuppertal, D-42119 Wuppertal, Germany}
\author{X.~Bai}
\affiliation{Bartol Research Institute and Department of Physics and Astronomy, University of Delaware, Newark, DE 19716, USA}
\author{M.~Baker}
\affiliation{Dept.~of Physics, University of Wisconsin, Madison, WI 53706, USA}
% ...

\date{\today}

\collaboration{IceCube Collaboration}
\noaffiliation

\begin{abstract}

A search for muon neutrinos from Kaluza-Klein dark matter annihilations in the Sun has been performed with the
22-string configuration of the IceCube neutrino detector using data collected in 104.3 days of live-time in 2007. No excess over the expected
atmospheric background has been observed. Upper limits have been obtained on
the annihilation rate of captured lightest Kaluza-Klein particle (LKP) WIMPs in the Sun and converted to limits on the LKP-proton cross-sections for LKP masses in the range 250 -- 3000 GeV.
These results are the most stringent limits to date on LKP annihilation in the~Sun.

\end{abstract}


\maketitle
In a recent work \cite{ic22_wimp}, we presented the result of a search for neutralino dark matter accumulated in the center of the Sun with the 22-string configuration of the IceCube detector. In this letter we extend the search to an alternative dark matter candidate, Kaluza-Klein (KK) particles, arising from theories with extra spacetime dimensions.
In the simplest framework of universal extra dimensions (UED) \cite{UED}, there is a single extra dimension of size $R \sim \mathcal{O} \quad \mathrm({TeV}^{-1})$ compactified on an $S^{1}/Z_{2}$ orbifold. Within minimal UED theories, the first excitation of the hypercharge gauge boson, $B^{(1)}$, is generally the lightest KK particle (LKP). It is often denoted as the KK 'photon', $\gamma^{(1)}$, since the effective first KK-level Weinberg angle of the mass matrix is very small, and therefore $B^{(1)}$ can also be described as a mass eigenstate \cite{UED}. KK-parity conservation, affiliated with extra-dimensional momentum conservation, leads to the stability of the LKP, which makes it a viable DM candidate. There are also other possible natural choices for LKP candidates within UED, like the KK 'graviton', the KK 'neutrino' or the $Z^{(1)}$-boson that may constitute viable DM candidates. They are not considered here. Instead, we focus on the most promising KKDM prospect in terms of indirect detection expectations, the KK 'photon'.\\
Accelerator measurements constrain the lower bound for the mass of the LKP, $m_{\gamma^{(1)}}$, at $300$ GeV \cite{acceleratorBound}. The upper bound is limited to a 
few TeV in order to not exceed the observed DM relic density and overclose the Universe.\\
We here consider UED models with five spacetime dimensions characterized by two parameters: the LKP mass, $m_{\gamma^{(1)}}$, and the mass splitting $\Delta_{q^{(1)}}\equiv (m_{q^{(1)}}-m_{\gamma^{(1)}})/m_{\gamma^{(1)}}$, where $m_{q^{(1)}}$ is the mass of the first KK quark excitation, as discussed in \cite{UED, servant, arrenberg, cheng}.
\\
As a possible dark matter component of the halo, LKPs can become gravitationally trapped in massive celestial bodies like the Sun, accumulating to high DM densities that can exceed the mean galactic density by several orders of magnitude in the object's core. Since the LKP is a boson, pair-wise annihilation is dominated by s-wave processes, creating standard model particles whose decay chains produce neutrinos in the GeV -- TeV range. The branching ratios for the LKP annihilation channels of interest are given in Table \ref{tab:br} for two values of $\Delta_{q^{(1)}}$ \cite{hooper}. The neutrinos may escape the Sun and reach Earth. The search presented here aims at detecting LKP annihilations indirectly by observing an excess of such high energy neutrinos from the direction of the Sun. Despite the existence of various limits on neutralino induced neutrino fluxes from the Sun \cite{baksan, macro, superk, sunWimp, ic22_wimp}, no corresponding limits for LKP annihilations have been previously 
reported.\par

\begin{table}
\caption{\label{tab:br} LKP annihilation branching ratios for two values of $\Delta_{q^{(1)}}$ \cite{hooper}. Ratios are not summed over generations. Channels within parenthesis give negligible contribution to a neutrino flux from the Sun. The Higgs-field annihilation channel, marked with $^{\dag}$, is neglected, due to large uncertainty and small contribution to the neutrino flux.}
\begin{ruledtabular}
 \begin{tabular}{c|cc}
 Channel & \multicolumn{2}{c}{Branching ratio}\\
         & $\Delta_{q^{(1)}}=0$&$\Delta_{q^{(1)}}=0.14$\\ \hline
$(e^{+}e^{-}),(\mu^{+}\mu^{-}),\tau^{+}\tau^{-}$                                            & 0.20 & 0.23\\
$(u\overline{u}),c\overline{c},t\overline{t}$                                               & 0.11 & 0.077\\
$(d\overline{d}),(s\overline{s}),b\overline{b}$                                             & 0.007 & 0.005\\
$\nu_{e}\overline{\nu}_{e},\nu_{\mu}\overline{\nu}_{\mu},\nu_{\tau}\overline{\nu}_{\tau}$   & 0.012 & 0.014\\
$(\Phi,\Phi^{\ast})^{\dag}$                                                                          & 0.023 & 0.027\\
\end{tabular}
\end{ruledtabular}
\end{table}

\begin{figure}[t]
  \centering
  \includegraphics[width=0.5\textwidth]{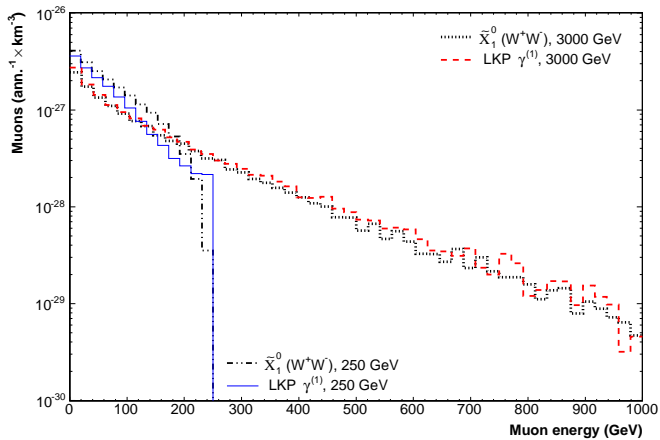}
  \caption{Comparison of simulated muon spectra from LKP, $\gamma^{(1)}$, and neutralino, $\tilde{\chi}^{0}_{1}$, annihilations observed in IceCube, for two WIMP masses, $250$ and $3000$ GeV, representing the boundries of the investigated LKP model space.
    }
  \label{fig:spectra}
\end{figure}

For the results presented here, we use the same data set, 104.3 days livetime taken with the 22-string configuration of IceCube in 2007, and the same analysis cuts as presented in~\cite{ic22_wimp}. This is justified since the signature of the expected signal at the detector is very similar for the LKP and neutralinos, considering the hardest $\tilde{\chi}^{0}_{1}$-annihilation channel into $W^{+}W^{-}$. The neutrino spectrum from annihilations of a LKP of a given mass in the center of the Sun is considerably harder than that of a neutralino of the same mass. However, oscillations and energy losses of the neutrinos on their way out of the Sun, like neutral current (NC) scattering, absorption and $\nu_{\tau}$-regeneration, smear out the energy spectra in a way that makes them comparable at Earth. Figure~\ref{fig:spectra} shows an example of how the resulting muon spectra at the detector compare for a selected choice of neutralino and LKP masses at $250$ and $3000$ GeV. The analysis strategy used in~\cite{ic22_wimp} is therefore already optimized for the search of KK dark matter.\par
We simulated LKP annihilations in the Sun using \texttt{WimpSim}~\cite{blen} for LKP masses $m_{\gamma^{(1)}}=250, 500, 700, 900, 1100, 1500, 3000$ GeV. We used $\Delta_{q^{(1)}}=0$ with annihilation branching ratios from Table~\ref{tab:br}. Since $\Delta_{q^{(1)}}>0$ results in an increased neutrino flux due to the importance of the contributions from the $\tau^{+}\tau^{-}$ and the direct neutrino channels, the choice of $\Delta_{q^{(1)}}=0$ leads to a conservative limit. The background in the search for neutrinos from the Sun comes from air showers created in cosmic ray interactions in the atmosphere. The showers cause downward going atmospheric muon events, triggering the detector at $\sim500$ Hz, and atmospheric muon neutrino events, triggering at $\sim4$ mHz. When the Sun is below the horizon, the neutrino signal can be distinguished from the atmospheric muon background by selecting events with upward--going reconstructed tracks.

\begin{table*}[t]
\caption{\label{tab:table2}Upper limits on the number of signal events $\mu_{\mathrm{s}}$, the conversion rate $\Gamma_{\nu\rightarrow\mu}$, the LKP annihilation rate in the Sun $\Gamma_{\mathrm{A}}$, the muon flux $\Phi_{\mu}$, and the LKP-proton scattering cross-sections (spin-independent, $\sigma^{\mathrm{SI}}$, and spin-dependent, $\sigma^{\mathrm{SD}}$), at the 90\% confidence level including systematic errors. The sensitivity $\overline{\Phi}_{\mu}$ (see text) is shown for comparison. Also shown is the median angular error $\Theta$, the mean muon energy $\langle\!E_{\mu}\!\rangle$, the effective volume $V_{\mathrm{eff}}$, and the $\nu_{\mu}$ effective area $A_{\mathrm{eff}}$.}
\begin{ruledtabular}
 \begin{tabular}{c|cccc|c|cc|cccc}
 $m_{\gamma^{(1)}}$&$\mu_{\mathrm{s}}$& $\Gamma_{\nu\rightarrow\mu}$&$\Gamma_{\mathrm{A}}$&$\Phi_{\mu}$&$\overline{\Phi}_{\mu}$&$\sigma^{\mathrm{SI}}$&$\sigma^{\mathrm{SD}}$&$\Theta$&$\langle\!E_{\mu}\!\rangle$&$V_{\mathrm{eff}}$&$A_{\mathrm{eff}}$\\
 (GeV)&  & ($\mathrm{km}^{-3} \mathrm{y}^{-1}$)& ($s^{-1}$) &($\mathrm{km}^{-2} \mathrm{y}^{-1}$)& ($\mathrm{km}^{-2} \mathrm{y}^{-1}$)&($\mathrm{cm}^{2}$)&($\mathrm{cm}^{2}$)&($^{\circ}$) &(GeV) & ($\mathrm{km}^{3}$)& ($\mathrm{m}^{2}$)\\ \hline
250 &  7.2 & $3.3\cdot 10^{3}$& $7.9\cdot 10^{21}$& $8.7 \cdot 10^{2}$& $1.7\cdot 10^{3}$& $4.9\cdot 10^{-43}$ & $3.7\cdot 10^{-40}$&$3.2$ & 65.8  & $7.6\cdot 10^{-3}$&$1.1\cdot10^{-4}$\\
500 &  6.9 & $1.2\cdot 10^{3}$& $2.2\cdot 10^{21}$& $4.6 \cdot 10^{2}$& $8.8\cdot 10^{2}$& $4.1\cdot 10^{-43}$ & $4.1\cdot 10^{-40}$&$3.0$ & 103  & $2.1\cdot 10^{-2}$&$4.0\cdot10^{-4}$\\
700&  7.3 & $9.2\cdot 10^{2}$& $1.7\cdot 10^{21}$& $4.1 \cdot 10^{2}$& $7.1\cdot 10^{2}$& $5.6\cdot 10^{-43}$ & $6.2\cdot 10^{-40}$&$2.9$ & 122  & $2.8\cdot 10^{-2}$&$5.7\cdot10^{-4}$\\
900&  7.0 & $7.8\cdot 10^{2}$& $1.5\cdot 10^{21}$& $3.8 \cdot 10^{2}$& $6.6\cdot 10^{2}$& $7.3\cdot 10^{-43}$ & $8.6\cdot 10^{-40}$&$2.9$ & 134  & $3.2\cdot 10^{-2}$&$6.6\cdot10^{-4}$\\
1100&  7.2 & $7.6\cdot 10^{2}$& $1.5\cdot 10^{21}$& $3.8 \cdot 10^{2}$& $6.6\cdot 10^{2}$& $1.0\cdot 10^{-42}$ & $1.3\cdot 10^{-39}$&$2.9$ & 141  & $3.3\cdot 10^{-2}$&$7.0\cdot10^{-4}$\\
1500&  7.2 & $6.6\cdot 10^{2}$& $1.3\cdot 10^{21}$& $3.5 \cdot 10^{2}$& $6.0\cdot 10^{2}$& $1.7\cdot 10^{-42}$ & $2.2\cdot 10^{-39}$&$2.9$ & 151  & $3.8\cdot 10^{-2}$&$7.9\cdot10^{-4}$\\
3000&  6.7 & $6.2\cdot 10^{2}$& $1.5\cdot 10^{21}$& $3.3 \cdot 10^{2}$& $5.8\cdot 10^{2}$& $7.4\cdot 10^{-42}$ & $9.9\cdot 10^{-39}$&$2.8$ & 152  & $3.8\cdot 10^{-2}$&$7.0\cdot10^{-4}$\\
\end{tabular}
\end{ruledtabular}
\end{table*} 
%
%
\begin{figure}[t]
  \centering
  \includegraphics[width=0.5\textwidth]{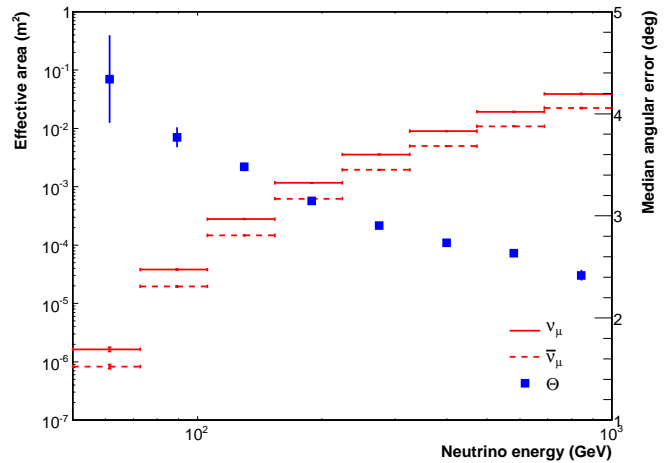}
  \caption{Lines showing the effective area (left scale) for the final event selection as function of neutrino energy in the range 50-1000 GeV, for muon neutrinos (solid line) and antineutrinos (dashed line) from the direction of the Sun. The result is an average over the austral winter. Systematic effects are included at the $1\sigma$ level, and statistical uncertainty of the same level are shown with error bars. Also shown is the median angular error $\Theta$ with $1\sigma$ error bars (squares, right scale).
    }
  \label{fig:aeff}
\end{figure}

Atmospheric muon and neutrino background events were also generated~\cite{cors,anis}. The propagation of muons and photons in the ice was simulated~\cite{mmc,pho} taking measured ice properties into account~\cite{iceprop}. \par
The events had to pass several selection criteria as described in~\cite{ic22_wimp} in order to reduce the content of atmospheric muon events. As a compromise between signal efficiency and background rejection, it was required that more than half of the events in the final data sample were neutrino-induced. The observables used describe the quality of the track reconstructions and the geometry and time evolution of the hit pattern in the detector, and they were required to be well reproduced in simulations. The event selection consisted first in a series of unidimensional cuts on the selected event variables, and a final step that used two Support Vector Machines (SVM). The SVMs were trained with simulated signal, and a set of experimental data, recorded in December 2007 and not used in this analysis since the Sun was above the horizon, was taken as background. A final sample was then defined from a cut on the combined two SVM output values, $Q_{1} \times Q_{2}$ (see figure 1 in~\cite{ic22_wimp}). The analysis was performed in a blind manner such that the azimuth of the Sun is unknown until the selection criteria were finalized.

\begin{figure}[t]
  \centering
  \includegraphics[width=0.5\textwidth]{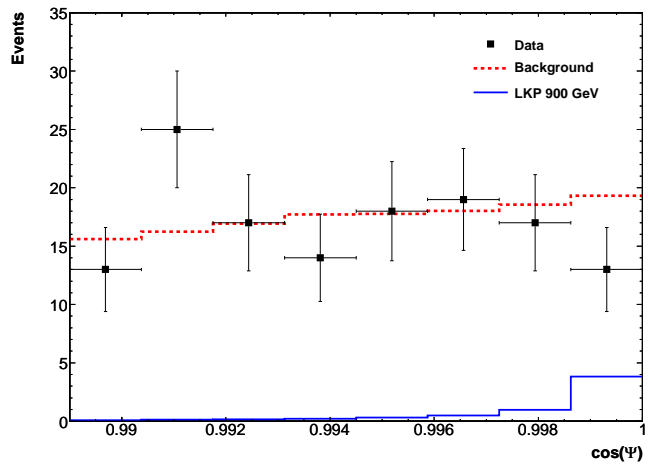}
  \caption{\label{fig:psi} Cosine of the angle between the reconstructed track and the direction of the Sun, $\Psi$, for data (squares) with one standard deviation error bars, and the atmospheric background expectation from atmospheric muons and neutrinos (dashed line). Also shown is a simulated signal ($m_{\gamma^{(1)}}$ = 900 GeV) scaled to $\mu_{s}=7.0$ events (see Table \ref{tab:table2}).}
\end{figure}

The systematic uncertainties on the effective volume, $V_{\mathrm{eff}}$, defined as the equivalent detector volume with $100\%$ selection efficiency, are the same as the ones calculated in the WIMP analysis in~\cite{ic22_wimp}, and are dominated by the uncertainties in photon propagation in the ice and the absolute DOM efficiency. They range from $\pm 19\%$ for the highest $m_{\gamma^{(1)}}$ to $\pm 26\%$ for the lowest $m_{\gamma^{(1)}}$ \cite{thesis}. From the final event selection of the signal simulation we additionally derive the effective area for muon neutrinos from the direction of the Sun as a function of neutrino energy, see Fig.~\ref{fig:aeff}. Also shown in the figure is the median angular error, the median of the angle between the reconstructed muon and the neutrino direction, $\Theta$. The result includes systematic uncertainties and is an average over the austral winter, during which the Sun is 
below the horizon.\par 
For the LKP signal models we then calculated the effective volume and, based on the distribution of the reconstructed angle to the Sun $\Psi$, we constructed confidence intervals at the $90\%$ confidence level using the method outlined in~\cite{ic22_wimp}: to evaluate the signal content in the final event sample, hypothesis testing was done based on $\Psi$, the angle between the reconstructed track and the direction of the Sun. From simulations we find $f_{\mathrm{s}}(\Psi)$, the probability distribution of $\Psi$ for the signal. By randomizing the azimuth angle in the final event sample of experimental data, $f_{\mathrm{b}}(\Psi)$, the equivalent probability distribution is found for background. Defining $\xi=\frac{\mu_{\mathrm{s}}}{n_{\mathrm{obs}}}$, from the number of signal events $\mu_{\mathrm{s}}$ and the observed number of events $n_{\mathrm{obs}}$, we form the combined probability density $f_{\xi}(\Psi)=\xi \cdot f_{\mathrm{s}}(\Psi)+(1-\xi)\cdot f_{\mathrm{b}}(\Psi)$. Based on $\xi_{\mathrm{best}}$, the non-negative signal content that maximizes the likelihood, we form the logarithm of the likelihood ratio $R(\xi)=\log(\prod_{i=1}^{i=n_{\mathrm{obs}}}\frac{f_{\xi}(\Psi_{i})}{f_{\xi_{\mathrm{best}}}(\Psi_{i})})$ \cite{fc}. Comparing this with a $R_{\mathrm{test}}(\xi)$ distribution of a large number of pseudo-experiments with $n_{\mathrm{obs}}$ events taken from $f_{\xi}(\Psi)$, we construct the confidence interval on $\xi$ at significance $\alpha$ as $R(\xi_{\it lim})=R_{\mathrm{test}}^{\alpha}(\xi_{\it lim})$, where $P(R_{\mathrm{test}} > R_{\mathrm{test}}^{\alpha})=1-\alpha$.\par

No excess of events from the Sun above the background expectation was found in the search $(\xi_{\mathrm{best}}=0)$. The observed number of events as a function of the angle to the Sun, $\Psi$, is compared to the atmospheric background expectation in Fig.~\ref{fig:psi}. From the upper limits on the number of signal events $\mu_{\mathrm{s}}$ we calculate the limit on the neutrino to muon conversion rate $\Gamma_{\nu\rightarrow\mu} = \frac{\mu_{\mathrm{s}}}{V_{\mathrm{eff}}\cdot t}$, for the livetime $t$. Using the signal simulation~\cite{blen}, we can convert this rate to a limit on the LKP annihilation rate in the Sun, $\Gamma_{\mathrm{A}}$, see Table~\ref{tab:table2}. Results from different experiments are commonly compared by calculating the limit on the muon flux above 1 GeV, $\Phi_{\mu}$, which is also given in Table~\ref{tab:table2} together with the sensitivity, $\overline{\Phi}_{\mu}$, the median limit obtained from simulations with no signal.\par

\begin{figure}[t!]
  \centering
  \includegraphics[width=0.5\textwidth]{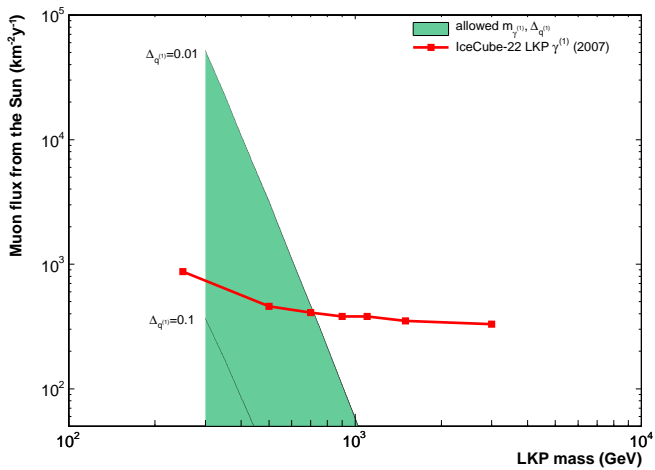}
  \caption{Limits on the muon flux from LKP annihilations in the Sun including systematic errors (squares), compared to the theoretically allowed region of $m_{\gamma^{(1)}}$ and $\Delta_{q^{(1)}}$. 
The regions corresponding to $\Delta_{q^{(1)}}=0.01$ and $\Delta_{q^{(1)}}=0.1$ are marked with black lines. The region below $m_{\gamma^{(1)}}=300$ GeV is excluded by collider experiments \cite{acceleratorBound}.}
  \label{fig:flux}
\end{figure}
\begin{figure}[t]
  \centering
  \includegraphics[width=0.5\textwidth]{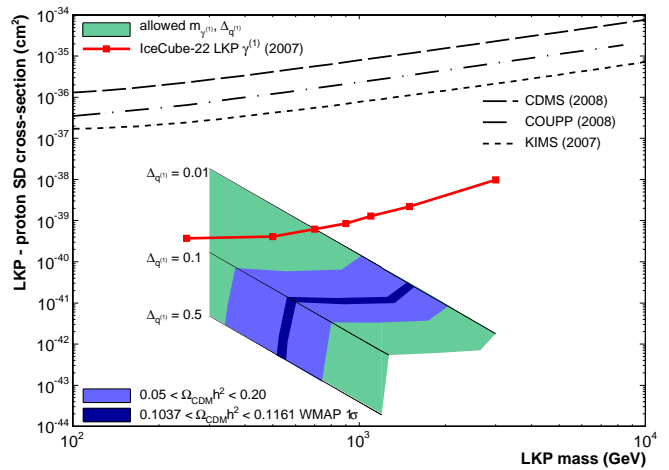}
  \caption{Limits on the LKP-proton SD scattering cross-section (squares) adjusted for systematic effects, compared with limits from direct detection experiments \cite{cdms,coupp,kims}. Theoretically predicted cross-sections are indicated by the green area \cite{arrenberg}. The regions corresponding to $\Delta_{q^{(1)}}=0.01,0.1,0.5$ are marked with black lines. The region below $m_{\gamma^{(1)}}=300$ GeV is excluded by collider experiments \cite{acceleratorBound} and the upper bound on $m_{\gamma^{(1)}}$ corresponds to the overclosure limit for each individual LKP model~\cite{servant}. The lighter blue region is allowed when considering $0.05<\Omega_{\mathrm{CDM}} h^{2}<0.20$, and the darker blue region corresponds to the preferred $1\sigma$ WMAP (5 year) relic density $0.1037<\Omega_{\mathrm{CDM}} h^{2}<0.1161$ \cite{dunkley}.
    }
  \label{fig:sd}
\end{figure}

The flux limit is shown in Fig.~\ref{fig:flux} together with the theoretically allowed flux region, derived from Refs.~\cite{arrenberg,hooper} with the use of DarkSUSY~\cite{dsusy}. We have here approximated the branching ratios for the regions of $\Delta_{q^{(1)}}=0.01$ and $\Delta_{q^{(1)}}=0.1$ with those of $\Delta_{q^{(1)}}=0$ and $\Delta_{q^{(1)}}=0.14$, respectively, as given in Table~\ref{tab:br}. The limits on the annihilation rate can be converted into limits on the spin-dependent, $\sigma^{\mathrm{SD}}$, and spin-independent, $\sigma^{\mathrm{SI}}$, LKP-proton cross-sections, allowing a comparison with the results of direct search experiments. Since capture in the Sun is dominated by $\sigma^{\mathrm{SD}}$, indirect searches are expected to be competitive in setting limits on this quantity. Assuming equilibrium between the capture and annihilation rates in the Sun, the annihilation rate is directly proportional to the cross-section. A conservative limit on $\sigma^{\mathrm{SD}}$ is found by setting $\sigma^{\mathrm{SI}}$ to zero, and vice versa. We have used the method described in Ref.~\cite{conv} to perform the conversion. The results are shown in Table~\ref{tab:table2}. We assumed a local WIMP density of $0.3~\mathrm{GeV/cm}^{3}$, and a Maxwellian WIMP velocity distribution with a dispersion of 270~km/s. Planetary effects on the capture were neglected~\cite{planet}. Figure~\ref{fig:sd} shows the limits on $\sigma^{\mathrm{SD}}$, as obtained with the
22-string configuration of IceCube compared with other bounds~\cite{cdms,coupp,kims}, and the KK model space. The theoretical model space (green area) is plotted for different predictions for the mass splitting $\Delta_{q^{(1)}}$. The blue regions indicate the overlap regions with two different $\Omega_{\mathrm{CDM}}$ intervals, whereas the narrow dark blue region corresponds to the preferred WMAP $1\sigma$-region for CDM. The upper bound on $m_{\gamma^{(1)}}$, derived from the overclosure limit for each individual LKP model~\cite{servant}, varies with different values of $\Delta_{q^{(1)}}$ and increases remarkably for models with $\Delta_{q^{(1)}}<0.1$. This is due to additional coannihilation effects, arising for degenerate LKP models~\cite{arrenberg}.

In conclusion, we have presented the first limits on LKP annihilations in the Sun. We also derived the most stringent limits on the spin-dependent LKP-proton cross sections in the non-excluded LKP mass regions ($300 \mathrm{GeV}<m_{\gamma^{(1)}}<3 \mathrm{TeV}$), improving existing limits by more than two orders of magnitude and excluding some viable LKP models. The full IceCube detector with the DeepCore extension \cite{dc} is expected to test most LKP models within the allowed region for $0.05<\Omega_{\mathrm{CDM}}h^{2}<0.20$, shown in Fig.~\ref{fig:sd}.\\

We thank the following agencies:
U.S. National Science Foundation-Office of Polar Programs,
U.S. National Science Foundation-Physics Division,
U. of Wisconsin Alumni Research Foundation,
U.S. Department of Energy, NERSC,
the LONI grid;
Swedish Research Council,
K.~\&~A.~Wallenberg Foundation, Sweden;
German Ministry for Education and Research,
Deutsche Forschungsgemeinschaft;
Fund for Scientific Research,
IWT-Flanders,
BELSPO, Belgium;
the Netherlands Organisation for Scientific Research;
M.~Ribordy is supported by SNF (Switzerland);
A.~Kappes and A.~Gro{\ss} are supported by the EU Marie Curie OIF Program.
We thank Sebastian Arrenberg and Kyoungchul Kong for helpful correspondence and details on their paper.

\end{document}